# Radiation from Comoving Poynting Flux Acceleration

*Edison Liang[1] and Koichi Noguchi[1]*

ABSTRACT

We derive analytic formulas for the radiation power output when electrons are accelerated by a relativistic comoving kinetic Poynting flux, and validate these analytic results with Particle-In-Cell simulations. We also derive analytically the critical frequency of the radiation spectrum. Potential astrophysical applications of these results are discussed. A quantitative model of gamma-ray bursts based on the breakout of kinetic Poynting flux is presented.

*Subject Heading*s: Acceleration of particles – Radiation mechanisms: non-thermal - Gamma-rays:bursts

*Online Material*: color figures

## 1. INTRODUCTION

In popular paradigms of radiation from blazars, pulsar wind nebulae (PWN), and gamma-ray bursters (GRB), relativistic outflow energy (hydrodynamic or electromagnetic) from the central compact object (black hole or neutron star) is first converted into relativistic nonthermal kinetic energy of electrons via some dissipation mechanisms (e.g. collisionless shocks, Dermer 2003, Meszaros 2002, Lyubarski 2005). These nonthermal electrons are then hypothesized to radiate, in the comoving frame of the outflow, synchrotron-like radiation (Rybicki and Lightman 1979, Epstein and Petrosian 1973, Lloyd and Petrosian 2000), or "jitter" radiation if the magnetic field is too chaotic (Weibel 1958, Medvedev 2000, Medvedev et al 2005). In addition, inverse Comptonization of the internal synchrotron (SSC) or external soft photons (EC), plus hadronic processes, may produce the high-energy gamma-rays (Dermer et al 2000, 2003). However, the kinetic processes which convert the outflow energy into nonthermal electron energy and radiation (Hoshino et al 1992, Gallant et al 1992, Silva et al 2003, Nishikawa et al 2003, Spitkovski 2006, Smolsky & Usov 2000, Lyutikov & Blackman 2002, Van Putten & Levinson 2003, Lyutikov and Blanford 2003) remain unsolved. In this paper we present a quantitative example of particle acceleration by a comoving Poynting flux (CPF), in which both the radiation power output and critical frequency can be derived analytically. We show that in this case the intrinsic radiation efficiency is very low compared to classical synchrotron theory in a static field. As a result electrons can be accelerated to very high Lorentz factors before radiation damping sets in.

In addition to the analytic theory, we have performed multi-dimensional Particle-in-Cell (PIC) simulations (Langdon and Lasinksi 1976, Birdsall & Langdon 1991, Langdon 1992) to model the nonthermal electron acceleration and radiation processes (Liang et al 2003, Liang & Nishimura 2004, Nishimura et al 2003, Liang & Noguchi 2005, 2006). A unique feature of our PIC simulations is that the intrinsic power radiated by each superparticle (=numerical representation of a charged particle) can be computed simultaneously as the superparticle is accelerated by the local Lorentz force (Noguchi et al 2005, Liang and Noguchi 2005, 2006). Such simulation provides a fully *self-consistent treatment of the intrinsic radiation power during the acceleration process*. We will calibrate and validate our analytic results using the PIC simulations. Section 2 reviews the basic physics of comoving PF acceleration (CPFA, this term replaces the acronyms DRPA and TPA used in our early papers). In Section 3 we derive the analytic formula for the radiation power output. Section 4 compares the analytic results with the numerical radiation power from PIC simulations. In Section 5 we derive analytically the critical frequency of CPFA radiation. In Section 6 we discuss the astrophysical applications of the above results.  In Section 7 we apply the analytic formulas to a simplified PF model of long GRBs.  Section 8 is devoted to discussions and summary.

## 2. COMOVING POYNTING FLUX ACCELERATION (CPFA)

In this paper we define "Poynting flux" (PF) narrowly as a kinetic plasma outflow dominated and accelerated by *transverse* electromagnetic (EM) fields with $\Omega_e/\omega_{pe}=B/(4\pi nm)^{1/2}$ >1, without the presence of flow-aligned guiding magnetic fields ($\Omega_e$ = eB/m =electron gyrofrequency, $\omega_{pe}=(4\pi ne^2/m)^{1/2}$=electron plasma frequency, m=electron mass, n=electron density, we set *c = 1 throughout this paper except in Sec.6 and Sec.7*).  Hence particle acceleration by classical Alfven and whistler waves (Boyd and Sanderson 1969, see discussions in Sec.8) in a background magnetic field, or by longitudinal plasma (Langmuir) waves (Tajima and Dawson 1979) will not be considered in this paper.  Instead we focus on semi-coherent particle acceleration by the ponderomotive (**JxB**) force of a comoving PF (Liang et al 2003). Astrophysical examples of such relativistic PF include the equatorial stripe wind of pulsars and magnetars (Lyubarsky 2005, Skjaeraasen et al 2005), and the low-density limit of a magnetic tower jet driven by strongly magnetized accretion disks around black holes (Koide et al 2004).  More examples will be discussed in Secs. 6 - 7.

Comoving PF acceleration (CPFA) occurs when an intense EM pulse, loaded with a small amount of plasma, maintains a *group* velocity (<c due to plasma loading) roughly in phase with the fastest electrons.  As slower electrons gradually fall behind the EM pulse, the plasma loading of the main EM pulse decreases, the pulse group velocity accelerates, and the Lorentz factor of the remaining comoving electrons increases, until dephasing or radiation damping sets in eventually.  The net effect is that the PF transfers its energy and momentum to a *decreasing* number of faster electrons over time (Liang & Nishimura 2004, LN04 hereafter).  A physical realization of CPFA was discovered by Liang et al (2003) using PIC simulations. When a static strongly magnetized ($B/(4\pi nm)^{1/2}>1$) overdense ($\omega_{pe}>2\pi/\lambda$, $\lambda$=characteristic wavelength of the EM pulse) plasma expands into a vacuum or low density region, the initial expansion disrupts the sustaining current, leading to $4\pi\mathbf{J} < \mathbf{Curl\ B}$. The excess displacement current ($\partial\mathbf{E}/\partial t$) then generates a transverse EM pulse, which tries to escape from the embedding plasma. As the EM pulse tries to escape, it "pulls" out the surface electrons via the $\mathbf{JxB}$ force (Fig.1a), where $\mathbf{J}$ is the self-induced polarization current (Boyd and Sanderson 1969). When the $\mathbf{JxB}$ force is very strong, the accelerated electrons can stay *comoving* with the *group* velocity of the EM pulse, and the acceleration becomes semi-coherent and self-sustaining (Liang et al 2003, Liang & Nishimura 2004=LN04, Liang 2005). CPFA can also be understood in terms of relativistic $\mathbf{ExB}$ drift in an intense EM pulse. As the drift velocity $v_d$ approaches c, the electron moves almost along a straight line (Fig.1b). Provided the plasma loading decreases with time due to the loss of slow particles, the EM pulse will accelerate, approaching a vacuum EM wave as |E/B| increases towards unity. This leads to continuous acceleration of $v_d$.

Using PIC simulations, LN04 found that the maximum Lorentz factor achievable by CPFA grows without limit as $\sim(\Omega_e t)^{1/2}$ until radiation damping or dephasing (e.g. due to wave-front curvature) sets in. LN04 also found that the CPF accelerates the high-energy electrons into a simple power law of slope –3 to – 4, independent of the initial conditions or the pulse size (Fig.2). CPFA is exceedingly robust and efficient, capable of converting >50% of the EM energy into accelerated particle energy over a distance ~ a few times the initial pulse width (see Sec.7). In contrast to shocks (Spitkovski 2006, Silva et al 2003), in which the bulk flow energy is converted into turbulent EM energy and internal particle heat, CPFA converts ordered EM energy directly into accelerated particle energy via continuous rarefaction of the plasma density.

The detailed physics of CPFA has been reviewed extensively elsewhere (Liang 2005, Liang and Noguchi 2005), so they will not be repeated here.

## 3. RADIATION POWER EMITTED BY CPFA

In this section we derive an analytic approximation for the power radiated by an electron accelerated kinetically by a CPF. The following derivation assumes linearly polarized plane EM waves for simplicity, but the results should be valid in general 3D geometry as long as the wave front curvature and transverse gradients are << 1/(acceleration distance). We emphasize that the radiation formula derived in this section should be applicable to any particle accelerated by transverse EM fields comoving with the local **ExB** drift velocity $v_d$. Hence its potential astrophysical applications should be much broader than the simplified CPFA scenarios discussed above.

Since the radiation power output (energy/sec) is a Lorentz invariant, one way to derive the power radiated is to start with the classical synchrotron formula (Rybicki and Lightman 1979) in a (primed) local Lorentz frame in which **E'** = 0 and **B'** is static, and then use appropriate Lorentz transformations to express the power in terms of lab-frame quantities. However, we find it to be more transparent to work directly in the lab-frame. It turns out to be also more convenient to discuss the different limiting cases and approximations if we derive the radiation power using lab-frame quantities. This is the approach we adopt in the following.

The relativistic dipole radiation power is given by (Rybicki and Lightman 1979):

$$P_{rad} = 2e^2 (F_{\|}^2 + \gamma^2 F_{+}^2)/3m^2 \tag{6}$$

where $\gamma$ = Lorentz factor, $F_{\|}$ = force component along velocity **v**, and $F_{+}$ = force component orthogonal to **v**. For particle motion in a linearly polarized plane EM wave with $(\mathbf{E},\mathbf{B}) = (E_z, B_y)$ (Fig.1b, note that $E_z$ is negative), we have the Lorentz force: $F_x=-ev_zB_y$; $F_y=0$; $F_z=e(E_z+v_xB_y)$. Here x is the direction of the Poynting vector **k**. After a little algebra we find:

$$F_\parallel=eE_zv_z/v; \quad F_+^2=e^2B_y^2[\sin^2\alpha(v^2-v_w^2)+(v_x-v_w)^2] \tag{2}$$

where $v_w=-E_z/B_y$ is the local "profile speed" of the EM field ($v_w<1$ due to plasma loading) and $\sin\alpha=v_z/v=p_z/p$. Substituting Eq.(2) into Eq.(1) we obtain:

$$P_{analytic} = 2e^4B_y^2[\sin^2\alpha(\gamma^2-1)(1-v_w^2)+\gamma^2(v_x-v_w)^2]/3m^2 \tag{3}$$

In addition to $B_y$ and $\gamma$, the instantaneous power radiated by an EM-wave-accelerated electron thus depends on two key parameters: the local EM field profile speed $v_w$ and the angle $\alpha$. Eq.(3) reduces to the classical synchrotron formula $P_{syn}=2e^4B_y^2\gamma p_+/3m^2$ ($p_+$ is the component of **p** orthogonal to **B**, Rybicki and Lightman 1979) in the static limit $v_w=0$ and simplifies in various other limits:

A. Comoving particles ($v_x=v_w$):

In this case Eq.(3) simplifies to $P_{analytic} = 2e^4B_y^2(p_z^2 + p_y^2)\sin^2\alpha/3m^2$ when $\gamma\gg1$. Since in all CPFA runs, $p_z\gg p_y$ at late times (see below), this reduces to

$$P_{analytic} = 2e^4B_y^2p_z^2\sin^2\alpha/3m^2 = 2e^4B_y^2p^2\sin^4\alpha/3m^2 \sim 2e^4B_y^2\gamma^2\sin^4\alpha/3m^2 \tag{4}$$

However, Eq.(4) is not a good approximation for electrons significantly out of phase with $v_w$ (note that electrons can have $v_x > v_w$ or $v_x < v_w$). Using PIC simulations, Liang and Nishimura (2004) showed that $v_w$ is closely related to the peak $\gamma$ of the particle distribution function $f(\gamma)$ of the main EM pulse (cf.Fig.2).

B. Vacuum pulse limit ($v_w=1$):

In the limit $v_w=1$, the PF propagates as a vacuum EM wave. Eq.(4) becomes, for $\gamma\gg1$:

$$P_{analytic} =2e^4B_y^2\gamma^2(1-v_x)^2/3m^2\sim e^4B_y^2\gamma^2\sin^4\alpha/6m^2 \tag{5}$$

Eq.(5) has the same functional form as Eq.(4) but its magnitude is a factor of 4 lower. It defines the *lower limit* to the radiative power loss of a PF-accelerated electron, since in practice $v_w<1$.

C. Slightly subluminal PF ($1-v_w=\varepsilon\ll 1$)

For most astrophysics applications, the PF will be slightly subluminal. We can simplify Eq.(3) by Taylor expanding $1-v_w=\varepsilon\ll 1$ to lowest order. This gives rise to:

$$P_{analytic} \sim 2e^4 B_y^2 \gamma^2 (\varepsilon + \sin^2\alpha/2)^2/3m^2 \tag{6}$$

For CPFA, both $\varepsilon$ and $\sin\alpha$ are $\ll 1$, leading to $P_{analytic} \ll P_{syn}$. This is because CPFA acts like a quasi-linear accelerator. Eq.(6) shows that $P_{analytic}$ behaves differently depending on whether $\varepsilon \gg$ or $\ll \sin^2\alpha/2$. In the former case $P_{analytic}$ depends only on the EM field profile speed $v_w$ and not on $\alpha$:

$$P_{analytic} \sim 2e^4 B_y^2 \gamma^2 \varepsilon^2/3m^2 \tag{7}$$

In the latter case we regain Eq.(5) which depends only on $\alpha$ and not on $v_w$. We emphasize that the comoving limit Eq.(4) is retrieved when $\varepsilon=\sin^2\alpha/2$. When we model astrophysical data using these formulas, we would obtain different $(B,\gamma)$ values depending on the values of $\varepsilon$ and $\sin\alpha$, which depends on the PF initial conditions. Eqs.(4), (5) & (7), which contain only 3 unknowns: $(B, \gamma, \alpha)$ or $(B, \gamma, \varepsilon)$, are easier to use for modeling astrophysical data than Eq.(3) or Eq.(6), which contain 4 unknowns. In all our CPFA simulations, $\sin\alpha$ lies in a narrower range for most high-power particles (Fig.6). So Eq.(6) is most useful in the ultra-relativistic regime $\varepsilon \ll \sin^2\alpha/2$, which seems to be the case for GRB's (cf. Sec.7) and may also be the case for blazars and PWNs. In the next section we will compare the above analytic approximations with PIC simulation results, which span a dynamic range of $10^5$.

4. NUMERICAL RADIATION POWER OUTPUT

In this section we present the intrinsic radiation output of electrons (and positrons) accelerated by a kinetic CPF using 2.5D (2d space, 3 momenta) PIC simulations and compare them to the analytic formulas of the last section. We compute the radiation power output by incorporating the relativistic dipole formula Eq.(1) into our PIC code. Numerically, we compute the power radiated by each superparticle by interpolating the Lorentz force data from the cell boundaries to the instantaneous superparticle position, so that **F** and **v** refer to the *same time and space* point (Noguchi et al 2005). This is a nontrivial procedure since in PIC simulations, particle and field data are offset by half time steps, and different field variables are offset by half grid spaces (Birdsall and Langdon 1991, Langdon and Lasinski 1976). We have carefully calibrated this numerical procedure with known analytic results. Fig.3 compares the PIC-simulated radiation output for an isotropic thermal plasma (kT=10m) in a static uniform B field, with that computed from the analytic synchrotron formula (Rybicki and Lightman 1979). Their good agreement, especially for the high-energy electrons, validates our numerical algorithm. However, PIC simulation data cannot be used to compute the radiation spectrum numerically because the PIC time step (typically = 0.25 gyroperiod) is too large to accommodate high frequencies in Fourier transforms.

Fig.4 highlights the time evolution of field and particle data of a typical CPFA run. A linearly polarized plane EM pulse accelerates a slab of overdense e+e- plasma from left to right, similar to the case studied by Liang and Nishimura (2004). While the energies of the pairs increase monotonically due to the CPFA (Fig.4a), the power radiated by the electrons rises to a maximum after ~ 5 light transit times of the pulse width, but then declines monotonically (Fig.4b) due to the trade-off between increasing γ and decreasing B and α. At late times (not shown) $P_{rad}$ approaches a constant value. In Fig.5 we compare the numerical $P_{rad}$ with $P_{analytic}$ of Eq.(4) for a different CPFA run. It shows good correlation for the highest-power particles, suggesting that these particles are comoving. At lower power the scatter-plot forms two separate bands above and below the 45º line, corresponding to electrons with $\varepsilon > \sin^2\alpha/2$ and $\varepsilon < \sin^2\alpha/2$ in Eq.(6) respectively. Fig.6 shows the distribution of $P_{rad}$ vs. $\sin\alpha$ for three CPFA runs of different initial temperatures. It shows that the highest-power particles have their $\sin\alpha$ values concentrated between 0.01 and 0.2. This narrow range of $\sin\alpha$ distribution seems to hold up in all our CPFA runs so far.

## 5. CRITICAL FREQUENCY OF CPFA RADIATION

A prominent feature of GRB and blazar spectra is the presence of a low energy spectral break $E_{pk}$ (hundreds of keV for classical GRBs, radio-IR for blazars). This spectral break is an indicator of the overall spectral hardness, and is usually interpreted as the critical frequency of synchrotron radiation $\omega_{crsyn} \sim 1.5\Omega_e\gamma_o p_+$ (Rybicki & Lightman 1979) by electrons with low energy cutoff $=\gamma_o$. This interpretation of the spectral break, plus the assumption of energy equipartition, are often used to constrain the Lorentz factor and magnetic field of the source. However, as we show below, for radiation emitted by CPFA electrons, the asymptotic critical frequency $\omega_{cr}$ is $<< \omega_{crsyn}$ due to the quasi-rectilinear motion.

To derive the formula for $\omega_{cr}$, we follow the approach of Landau and Lifshitz (1980): $\omega_{cr}$ is determined by the time measured in detector frame it takes the radiation beam of opening angle $1/\gamma$ to sweep past the detector due to the curvature of the particle trajectory. For electrons comoving or almost comoving with the PF, the parallel momentum $p_x$ (x is the direction along **k**, Fig.1a) increases monotonically while $p_z$ (momentum along **E**) asymptotes to a constant (Liang & Nishimura 2004, note that $p_y$ along **B** is constant to first order). Hence the change in the radiation beam direction due to bending of particle trajectory is dominated by the change in $p_x$:

$\Delta\theta \sim p_z\Delta p_x/p_x^2$. From the Lorentz force equation we have $d\gamma/dt = eE_zp_z/m\gamma$. Hence the time in the laboratory frame for the radiation beam to change by an angle $\Delta\theta \sim 2/\gamma$ is $\Delta t = 2\gamma^2 m/(eE_zp_z^2)$ where we have used the approximation $\gamma \sim p_x (>> p_z, p_y)$. This translates into a duration in the detector frame $\Delta t_{ob} = \Delta t/2\gamma^2 = m/eE_zp_z^2$. Thus the critical frequency (Rybicki and Lightman 1979):

$$\omega_{cr} = 1.5/\Delta t_{ob} = 1.5eE_zp_z^2/m = 1.5\Omega_e p_z^2 \sim 1.5\Omega_e\gamma^2\sin^2\alpha \sim \omega_{crsyn}\sin^2\alpha \qquad (8).$$

Since $\sin\alpha << 1$ at high power (Fig.6), $\omega_{cr} << \omega_{crsyn}$. In Sec.7 we will discuss the implications of this result for modeling GRB data. In Fig.4c we show the time evolution of $\omega_{cr}$ for the same CPFA run as Figs.4ab. It shows that spectral hardness follows the same trend as $P_{rad}$, reaching a maximum after a few light transit times of the pulse width. However the decline of $P_{rad}$ is more rapid than $\omega_{cr}$ due to the extra factors of $\sin\alpha$.

## 6. ASTROPHYSICAL APPLICATIONS

One of the most important applications of the above results to astrophysics is the extraction of the (B, γ) values of the emission region from observational data. Conventional synchrotron models leave (B,γ) as free parameters to be determined using other ad hoc assumptions such as energy equipartition, and injection rates of nonthermal electrons (Dermer and Boettcher 2002). For the CPFA model, however, we can tightly constrain (B,γ) from first principles because the electron acceleration rate is uniquely determined by the PF magnetic field. Liang and Nishimura (2004) derived, using the comoving Lorentz equation, the particle acceleration rate for CPFA electrons:

$$d\gamma/dt = f\Omega_e/\gamma, \qquad (9)$$

where f is a fudge parameter of O(1) that depends only weakly on the initial PF magnetization $\Omega_e/\omega_{pe}$. *This important result is independent of the global properties of the plasma or details of* *the EM pulse profile.* By equating Eq.(9) to the radiation loss rate of Sec.3, we can now express the limiting Lorentz factor $\gamma_{max}$ of CPFA electrons due to radiation damping:

$$\gamma_{max} = (3fm^2c^4/(2Be^3\sin^4\alpha))^{1/3} \qquad (10).$$

For a given B field, Eq.(10) predicts a Lorentz factor much higher than typical limits from synchrotron radiation damping since $\sin\alpha \ll 1$. Eq.(10), together with Eq.(8) for the critical frequency, allow us to determine $(B,\gamma)$ uniquely from the spectral break energy $E_{pk}$ (modulo the small uncertainty in f and $\sin\alpha$), without invoking energy equipartition, particle injection rate or other ad hoc assumptions. In this sense the *CPFA model is more constraining and predictive than conventional synchrotron models,* which do not specify the particle acceleration mechanism or provide the acceleration rate. In Sec.7 we will apply Eq.(10) and Eq.(8) to a PF model of GRBs.

CPFA may be relevant to astrophysics in two different types of settings: global and local. Globally, macroscopic EM pulses with ordered fields and low plasma loading may be generated by magnetic tower jets or transient magnetar wind (Koide et al 2004) emerging from collapsars, or from the merger of strongly magnetized neutron stars into a black hole. For example, CPFA can take place when a magnetic tower jet punches through a collapsar envelope or wind, and converts into a kinetic EM pulse at sufficiently low ambient density (see Sec.7). Similarly, when a millisecond magnetar collapses, or when a strongly magnetized neutron star binary merges, to form a black hole, part of its collapse energy may be emitted in the form of an intense EM pulse.

Alternatively, CPFA may also occur at the local level in the absence of large scale ordered EM fields. For example, magnetic-dominated (high-$\Sigma$) turbulence generated by relativistic shear layers, shocks or reconnection, may dissipate locally via the CPFA mechanism when nonlinear EM waves propagate into low density regions with $\Omega_e/\omega_{pe}>1$. In this case CPF acceleration persists only until dephasing occurs due to wave front curvature and inhomogeneity. So the maximum Lorentz factor achieved may be much lower than the radiation damping limit given by Eq.(10).

## 7. APPLICATION TO A PF MODEL OF LONG GRB's

Currently there is no universally accepted model of GRB energization and radiation. Two popular paradigms are hydrodynamic or electromagnetic outflows from a central engine (e.g. a newly formed black hole accretion disk or millisecond magnetar), dissipating at a distance of $10^{14-15}$ cm (Meszaros 2002, Piran 2004). Recent Fermi observation of ultra-luminous GRB080916c which fits a simple "Band function" spectrum extending from soft x-rays to >100 GeV (Abdo et al 2009) may favor a PF origin (Zhang 2009). If GRB is indeed energized by an intense PF outflow, CPFA would be an attractive dissipation mechanism due to its high energy conversion efficiency (Fig.7) and universal power-law spectra with low energy turnover (Fig.2). Here we apply the analytic formulas of the previous sections to a simple model of long GRBs, assuming that the PF contains only e+e- pairs with no ions (e-ion models will be considered in future papers). *We can predict the spectral break energy $E_{pk}$ from first principles, which has not been achieved in conventional synchrotron shock models.*

Our underlying astrophysical picture is that some central engine activity lasting 10's of seconds launches an intense EM pulse of width ~ $10^{12}$ cm and energy ~$10^{51}$ ergs, loaded with low-density e+e- plasma so that $\Omega_e/\omega_{pe} >> 1$. This intense EM pulse initially propagates through a collapsar envelope as a non-relativitic MHD pulse since the ambient density is high and the formal Alfven speed $v_A = B/(4\pi\rho_p)^{1/2} << c$ ($\rho_p$ = ambient gas mass density). But the pulse eventually reaches a point where the ambient density is so low that $v_A \geq c$, and the MHD pulse "breaks out" into a kinetic EM expansion similar to the scenario studied by Liang and Nishimura (2004). This PF "breakout" triggers the CPFA and rapid conversion of EM energy into particle energy. We have performed PIC simulations of relativistic magnetosonic pulses propagating down steep density gradients. The preliminary results seem to support the above "breakout" picture. We emphasize that in this case the GRB "ejecta" is just a bundle of EM energy with current-carrying leptons which act as radiation agents with little inertia.

For long GRBs, it is useful to scale the burst parameters with the following benchmark values (Fishman & Meegan 1998, Preece et al 2000): total energy $E_{51}=E_{tot}/10^{51}$erg, burst duration $T_{30}=T/30$sec, prompt-γ emission distance $R_{14}=R/10^{14}$cm. We approximate the EM pulse as a quasi-spherical shell with thickness $\Delta R=cT=10^{12}$cm$T_{30}$ (in this section we write out c explicitly) and solid angle $\Omega_{4\pi}=\Omega/4\pi$. To simplify the model we assume that the shell is uniform with mean field B and mean lepton (e- + e+) density n. *All physical quantities are measured in the "lab-frame", which we assume to be the rest frame of the GRB central engine*. In reality the field, density and particle momentum profiles are highly structured due to current instabilities (Liang and Nishimura 2004), and the following parameters primarily refer to those leptons at the peak of the momentum distribution function. PIC simulations suggest that at late times, particle energy $E_{particle} \sim 0.6E_{tot}$, EM energy (= $2E_B$)~ $0.4E_{tot}$ (Liang et al 2003). Let N = total number of leptons (e+ + e-) in the pulse and Γ = peak Lorentz factor of the lepton distribution = $\langle\gamma f(\gamma)\rangle$. (Fig.2, Γ ~ the group velocity Lorentz factor of EM pulse =$(1-v_w^2)^{-1/2}$, Liang and Nishimura 2004). In cgs units we have dimensionally:

$$N\Gamma mc^2 \sim 6\times10^{50} E_{51} \qquad (11)$$

$$B^2\Delta R R^2\Omega \sim 16\pi\times10^{50} E_{51} \qquad (12)$$

Eq.(12) gives:

$$B \sim 2\times10^5 \text{ G } (R_{14}^{-1} \Omega_{4\pi}^{-1/2} E_{51}^{1/2} T_{30}^{-1/2}) \qquad (13)$$

Next we estimate $\Gamma$ by invoking Eq.(10):

$$\Gamma \sim 1.2 \times 10^5 \, (f^{1/3} R_{14}^{1/3} \Omega_{4\pi}^{1/6} E_{51}^{-1/6} T_{30}^{1/6} \alpha_{.1}^{-4/3}) \tag{14}$$

where we have scaled $\sin\alpha$ with 0.1: $\alpha_{.1} = \sin\alpha/0.1$. Hence $\varepsilon \sim 1/\Gamma \ll \sin\alpha$ and our assumption of ignoring $\varepsilon$ in Eq.(6) is justified. Using this in Eq.(11) we find:

$$N \sim 6 \times 10^{51} \, (f^{-1/3} R_{14}^{-1/3} \Omega_{4\pi}^{-1/6} E_{51}^{7/6} T_{30}^{1/6} \alpha_{.1}^{4/3}) \tag{15}$$

Combining Eqs.(8), (13) and (14) we obtain the value of the spectral break energy, taken as the critical frequency corresponding to $\Gamma$:

$$E_{pk} = \hbar\omega_{cr}/2\pi \sim 490 \text{ keV}(f^{2/3} R_{14}^{-1/3} \Omega_{4\pi}^{-1/6} E_{51}^{1/6} T_{30}^{-1/6} \alpha_{.1}^{-2/3}) \tag{16}$$

Interestingly, this value, which is derived from first principles using the CPFA model, agrees with typical spectral break energies of long GRBs in the host-Galaxy frame: $E_{pk} \sim 250 \text{keV} (1+z) \sim 500$ keV for $z \sim 1$ (Preece et al 2000). Eq.(16) depends only weakly on the constituent parameters so the result is rather robust. It would be interesting to explore the implication of Eq.(16) for the Amati-Ghirlanda-type relations (Amati et al 2002, Ghirlanda et al 2004) by studying the dependence of T, $\alpha$, f etc on the total energy E. We emphasize that conventional synchrotron models of GRBs do not predict values of $E_{pk}$ from first principles since the acceleration mechanism is not specified.

From Eq.(15) we obtain the mean lepton density:

$$n = N/(\Omega \Delta R R^2) \sim 5 \times 10^{10} (f^{-1/3} R_{14}^{-7/3} \Omega_{4\pi}^{-1/6} E_{51}^{7/6} T_{30}^{-7/6} \alpha_{.1}^{4/3}) \tag{17}$$

and the magnetization:

$$\Omega_e/\omega_{pe} \sim 250 \, (f^{1/6} R_{14}^{1/6} \Omega_{4\pi}^{-5/12} E_{51}^{-1/12} T_{30}^{1/12} \alpha_{.1}^{-2/3}) \gg 1 \tag{18}$$

which justifies our EM-domination assumption. At this density the pairs are completely collisionless (Coulomb mean free path $> 10^{20}$cm). We note that the local acceleration time of an individual lepton, with the above values of B, $\Gamma$ and $\sin\alpha$, is very short: $t_{acc} = t_{rad} \sim 10^{-2}$ sec, which means that the leptons quickly achieve their asymptotic Lorentz factor $\Gamma$ after PF breakout. However, the overall cooling/dissipation time of the EM pulse is determined by the time to convert the global EM energy into lepton energy, which is proportional to the light transit time across the shell thickness $\Delta R/c$ (Fig.7). Moreover, radiation emitted by the front and back of the plasma pulse also arrive at the detector with a time delay of $\Delta R/c$. These two effects combine to make the GRB duration measured by the detector $\sim \Delta R/c = 30$ sec (Fig.7), irrespective of the short acceleration time of individual leptons. We note that $10^{12}$cm corresponds to $\sim 10^{14}$ gyroradii, and the acceleration length of $\sim 3\times 10^{8}$cm still equals $10^{10}$ gyroradii. Both scales are much larger than the largest PIC simulations we have performed ($\sim 10^{7}$ gyroradii). However, we emphasize that the only physics invoked to derive the acceleration and radiation cooling rates are all scale invariant. Fig.7 also demonstrates the scalability of the overall energy conversion rate. Hence we are reasonably confident that our kinetic results can be applied to macroscopic astrophysical systems.

However, one puzzle remains: why and how does the EM pulse decide to dissipate at $R \sim 10^{14}$ cm from the central engine, two orders of magnitude larger than the EM pulse width and six orders of magnitude larger than the lepton acceleration length? We speculate that it may be the GRB environment which determines this dissipation distance. Here we venture a speculative but plausible scenario that gives rise to such a far away dissipation site from the central engine. In reality, the PF "breakout" takes place not at a sharp star-vacuum boundary but in an external density gradient whose scale height is much larger than the pulse width or acceleration length. Hence we speculate that Eqs. (9) and (10) are valid only when the ambient ion mass density drops below the internal pair mass density. Otherwise the EM expansion and particle acceleration would be strongly inhibited by the ambient ion inertia. In the collapsar model, the GRB progenitor is likely surrounded by a Wolf-Rayet wind whose mass density $\sim A.5 \times 10^{11}$ $r^{-2}$ g.cm$^{-1}$ (Chevalier and Li 2000) where the parameter A depends on the mass loss rate. Hence the PF "breakout" distance, using the pair density of Eq.(17), becomes $r_{breakout} \sim A^{1/2} 10^{14}$ cm. In other words, the PF breakout and lepton acceleration are inhibited by ion inertia of the progenitor wind, until the PF reaches an ambient ion mass density of $\leq 5 \times 10^{-17}$ g.cm$^{-3}$, which only occurs at a distance $\geq A^{1/2} 10^{14}$ cm. Fig.8 illustrates the relevant scales discussed in this scenario.

## 8. DISCUSSIONS AND SUMMARY

We have shown in this paper that when electrons/pairs are accelerated by a comoving Poynting flux with $\Omega_e/\omega_{pe} > 1$, the intrinsic radiation power and critical frequency can be estimated analytically, and the values measured in the laboratory frame are much below those expected from synchrotron in a static field. This is because the EM field is almost comoving with the high energy particles and the particle paths are quasi-rectilinear. We apply our formulas to a simple PF "breakout" model of classical long GRBs, and find that the predicted spectral break energy agrees with the range of observed $E_{pk}$ values.

Besides the CPFA mechanism, there are many other Poynting flux scenarios that can lead to nonthermal particle acceleration. For example, electron acceleration by longitudinal wakefields generated by PF in an underdense plasma (similar to laser accelerators in the laboratory, Tajima and Dawson 1979) may occur in special astrophysical situations. We have also not considered Poynting flux dominated by Alfven and whistler waves. In general, waves of all types can transfer energy to electrons via resonant scatterings (Boyd and Sanderson 1969). But resonant interactions tend to only act on a small fraction of the electrons at any time, whereas the ponderomotive force of CPFA accelerate the bulk of the plasma in a sustained manner, and transfer most of the EM energy to particles. PF acceleration of e-ion plasmas is more complex than e+e- plasmas due to charge separation (Nishimura et al 2003). Their radiation output will be studied in a separate paper.

This work was partially supported by NSF AST0406882 and NASA NNG06GH06G.

1. Rice University, Houston TX 77005-1892.

Figure Captions

Fig.1 (a) Picture illustrating the CPFA concept. An intense plane EM pulse escaping from an overdense plasma induces a polarization current, such that the **JxB** force pulls out the surface electrons relativistically. But only the fastest electrons can keep up with the EM pulse. So the plasma loading of the EM pulse decreases with time. This leads to sustained acceleration of only the fast electrons, with no limit to their Lorentz factor. The sharp plasma boundary (hatched) is sketched only for illustration. Actual PIC simulations using smooth density profiles achieve similar asymptotic results. In all figures of this paper, x is expressed in units of $3c/\omega_{pe}$. (b) CPFA can also be visualized in terms of **ExB** drift in a comoving EM pulse. As $|E| \rightarrow |B|$, the particle path becomes quasi-rectilinear. $\alpha$ is the asymptotic angle between the Poynting vector **k** and the drift velocity $\mathbf{v_d}$.

Fig.2 (a) Electron energy spectra of two CPFA PIC runs with different initial pulse widths (a) $L_o=10^4$ $c/\omega_{pe}$ and (b) $L_o=10^3$ $c/\omega_{pe}$. Both develop robust power-laws of slope ~ -3 to –4 with low energy turnovers. Spectrum (a) was obtained after $10L_o/c$. Spectrum (b) was obtained after 100 $L_o/c$, and the pulse group Lorentz factor has reached $\gamma_w$ ~ 15 as evidenced by the turnover energy. It takes several million time steps to achieve such well-defined power laws.

Fig.3 Calibration of the numerical radiation power $P_{rad}$ computed from the PIC simulation (Eq.(1)) against the analytic synchrotron formula $P_{syn}$ for a 5 MeV thermal plasma in a static uniform B field shows good agreement in both (a) log-log and (b) linear-linear plots. The agreement is best for high power particles. The small scatter is due to numerical errors from interpolating the field values to the particle positions. In all figures of this paper, $P_{rad}$, $P_{syn}$ and $P_{analytic}$ are expressed in units of $2e^2\Omega_e^2/2700$.

Fig.4 Time-lapse progression of a pair-loaded EM pulse from left to right shows the CPFA in action. The e+e- plasma slab has initial temperature $kT_o=0.005m$, pulse width $L_o=12c/\omega_{pe}$, $\Omega_e/\omega_{pe}=10$, and was initially centered at $x\omega_{pe}/3c=180$. The 5 snapshots (left to right) are taken at $t\omega_{pe}/3=0, 12, 20, 60, 100$. We plot only 1% of all superparticles in the scatter-plots. (a) Progression of the magnetic field $B_y$ and electron Lorentz factors $\gamma$ distribution shows monotonic increase in $\gamma$ and conversion of magnetic energy into particle energy via current dissipation. (b) Evolution of $P_{rad}$ distribution shows that radiation loss for the highest energy electrons peaks at ~5 light crossing times, followed by monotonic decay. The increase in $\gamma$ is countered by the decrease in B and $\sin\alpha$. (c) Evolution of the critical frequency $\omega_{cr}$ distribution shows a similar trend as $P_{rad}$. $\omega_{cr}$ is expressed in units of $10\Omega_e$.

Fig.5 Scatter-plot of $P_{rad}$ versus $P_{analytic}$ of Eq.(4) for an e+e- CPFA run with initial $\Omega_e/\omega_{pe}=10$ and $kT_o = 10m$. We see that the two powers converge for the highest power particles. At lower power $P_{rad}$ deviates from Eq.(4). The scatter is concentrated in two bands above and below the 45° line. These bands correspond to electrons with $\varepsilon > \sin^2\alpha/2$ and $\varepsilon < \sin^2\alpha/2$ in Eq.(6) respectively.

Fig.6 Scatter-plots of the distribution of $P_{rad}$ vs. $\sin\alpha$ at late times for three sample CPFA runs with initial $\Omega_e/\omega_{pe}=10$ and $kT_o =$ (a) 10m; (b) 0.005m; (c) 0.125m. Horizontal dash lines denote $P_{rad} =1\%$ of the maximum emitted power. Most high power electrons have $0.01 \leq \sin\alpha \leq 0.2$ (vertical dotted lines). The maximum powers are emitted at angles $0.02 \leq \sin\alpha \leq 0.1$ in all cases.

Fig.7 Decay of EM energy for $\Omega_e/\omega_{pe}=10$, $kT_o=10m$ CPFA runs with three different initial pulse widths: (A) $L_o=10800c/\omega_{pe}$; (B) $L_o=90c/\omega_{pe}$; (C) $L_o=12c/\omega_{pe}$. These results show that the time to convert 50% of EM energy into particle energy is simply proportional to the light transit time $L_o/c$.

Fig.8 Schematic diagram illustrating the different physical scales in the "breakout" of a PF from a Wolf-Rayet wind model for long GRBs. The wavy arrow denotes the (lab-frame) PF pulse width ($\Delta R=10^{12}$cm) along the observer line of sight. The PF breakout distance ($\sim 10^{14}$ cm) is determined by the radius at which the ambient wind mass density drops below the PF internal pair mass density ($\sim 5 \times 10^{-17}$ g.cm$^{-3}$). Despite the short acceleration length ($\sim 3 \times 10^{8}$cm) of individual leptons as the pulse emerges, the detector-measured GRB duration at infinity is determined by the overall transit time $\sim \Delta R/c=30$ sec of the pulse passing through $r_{breakout}$ and the light paths between the front and back of the PF pulse (upper-right space-time diagram).

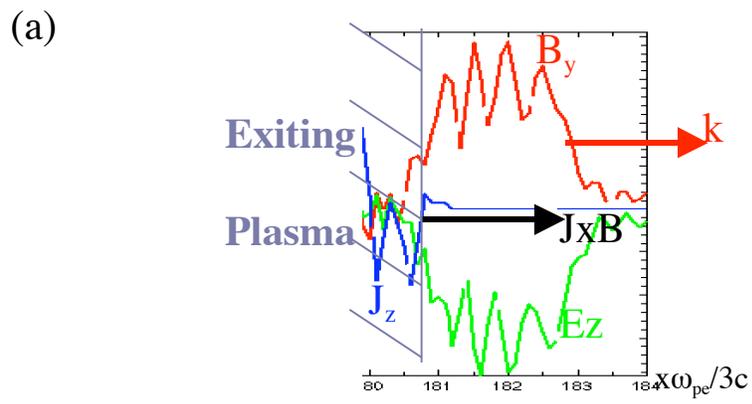

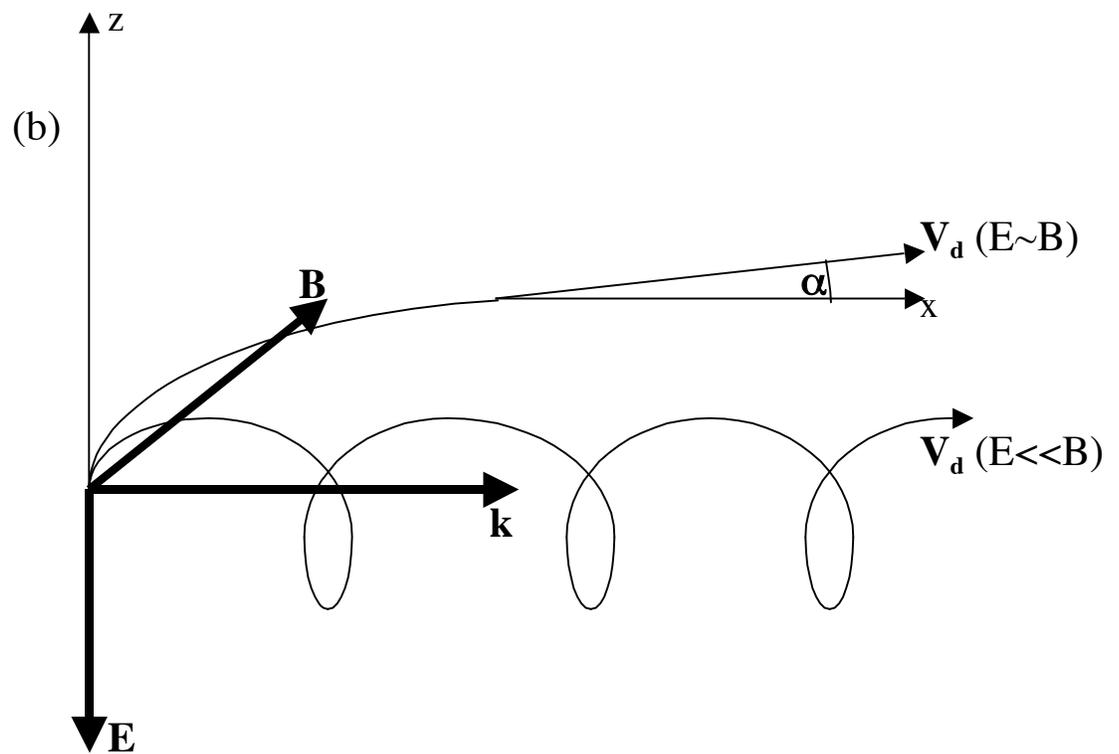

Fig.1

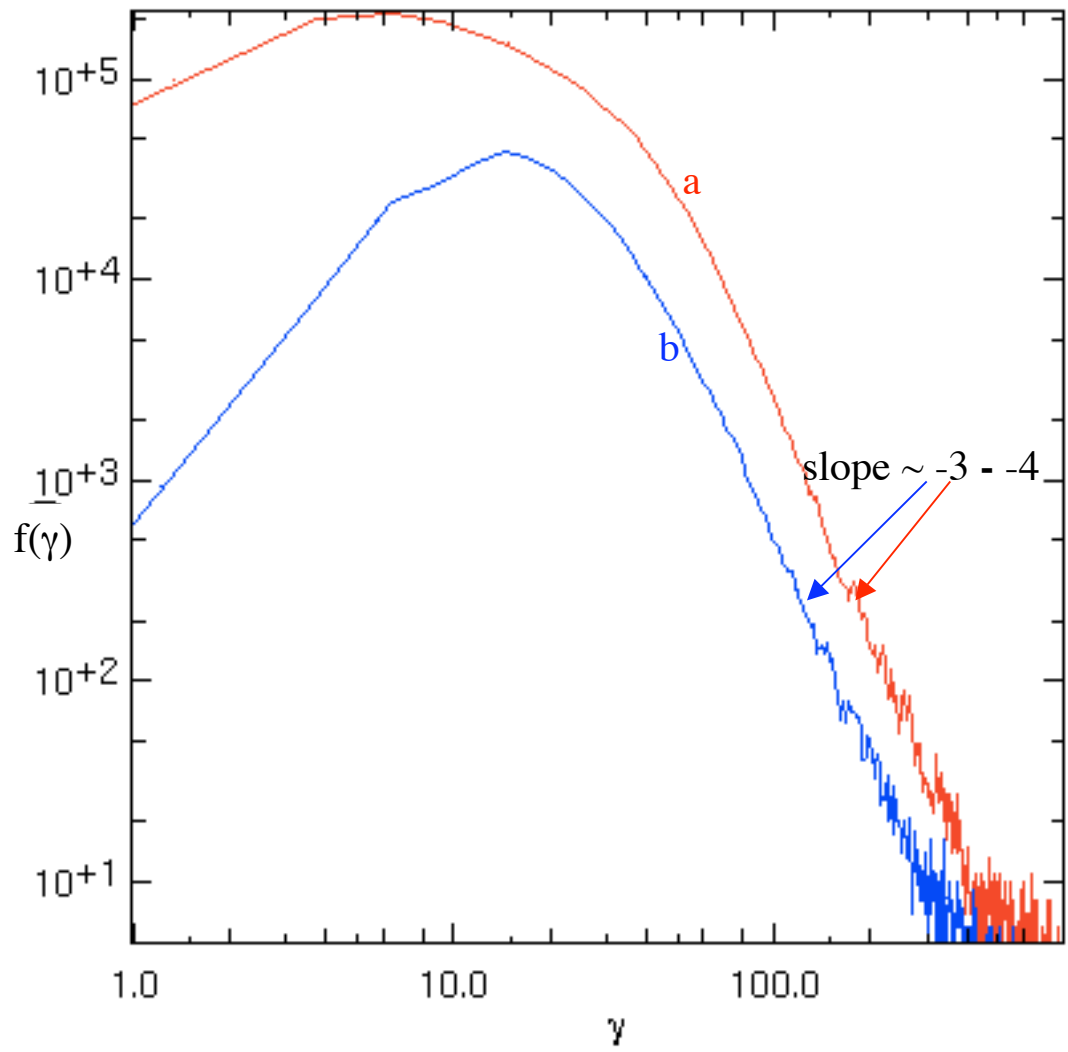

Fig.2

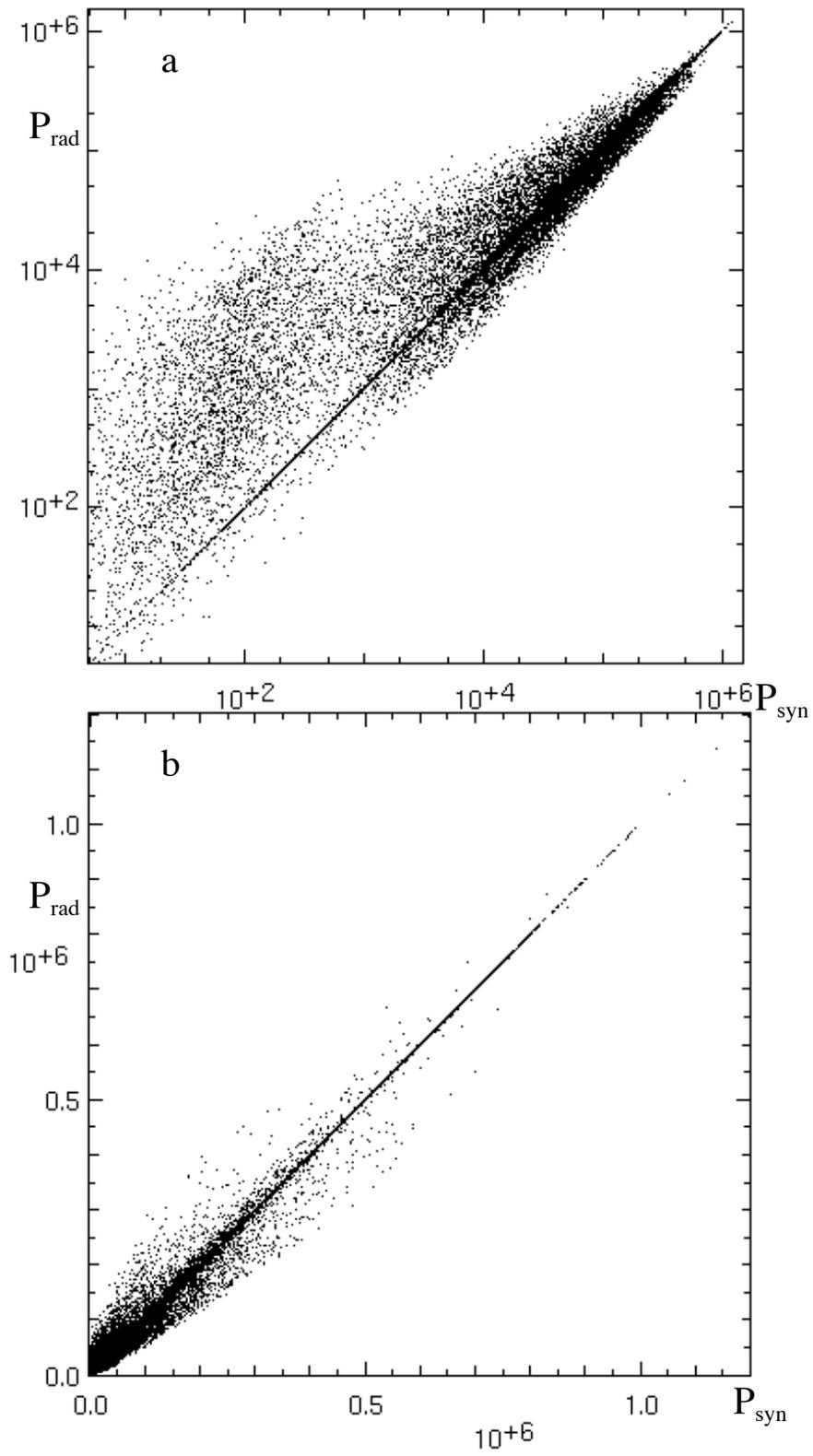

Fig.3

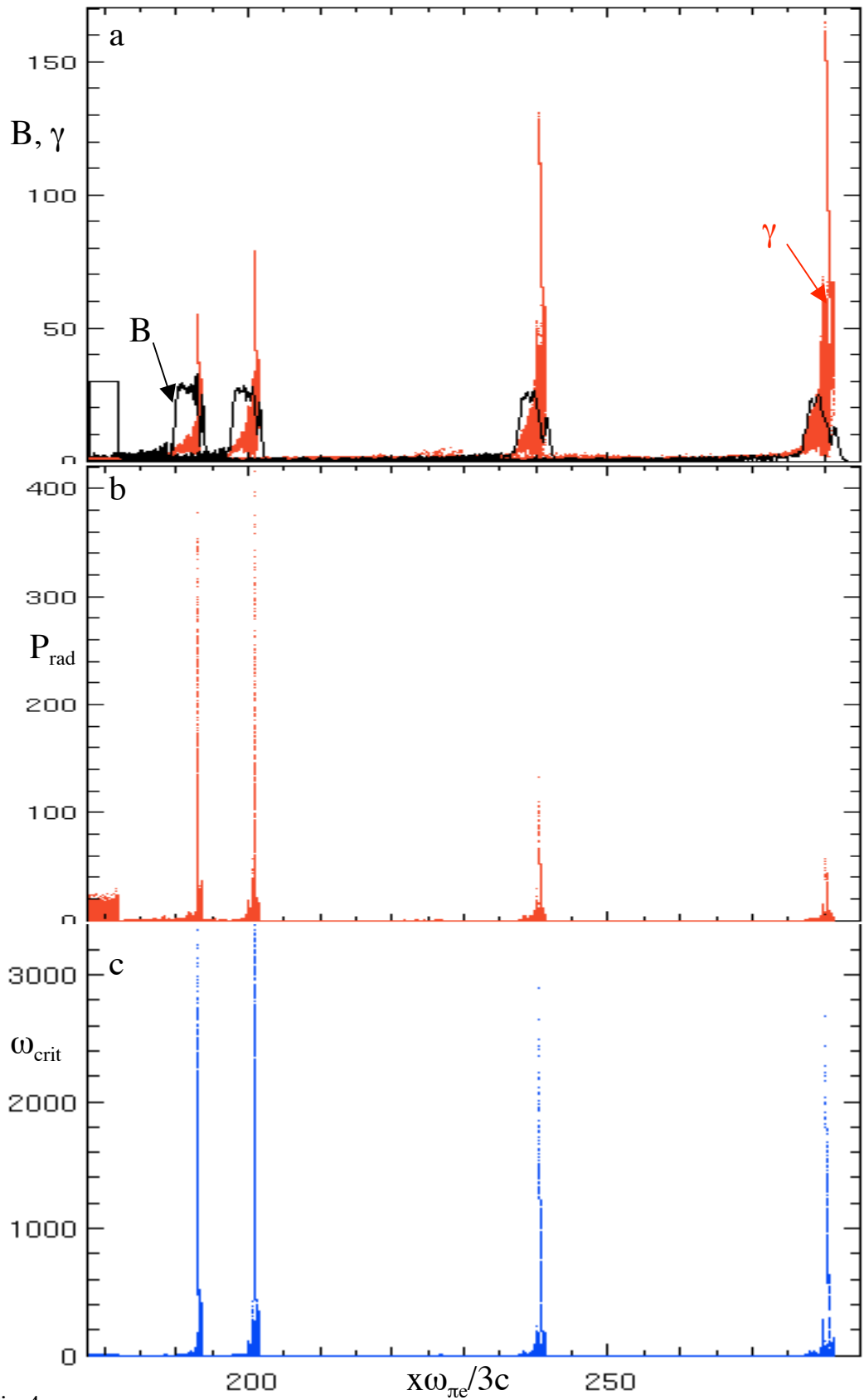

Fig.4

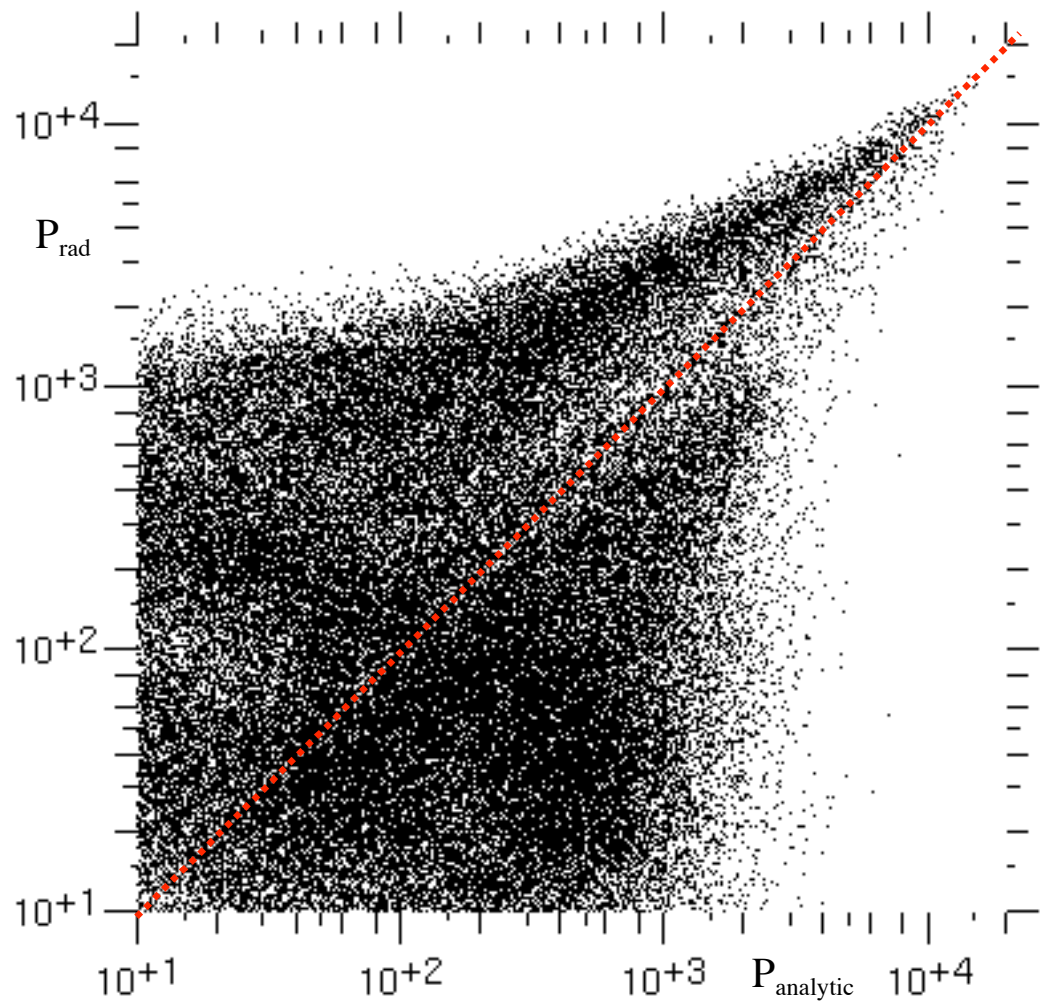

Fig.5

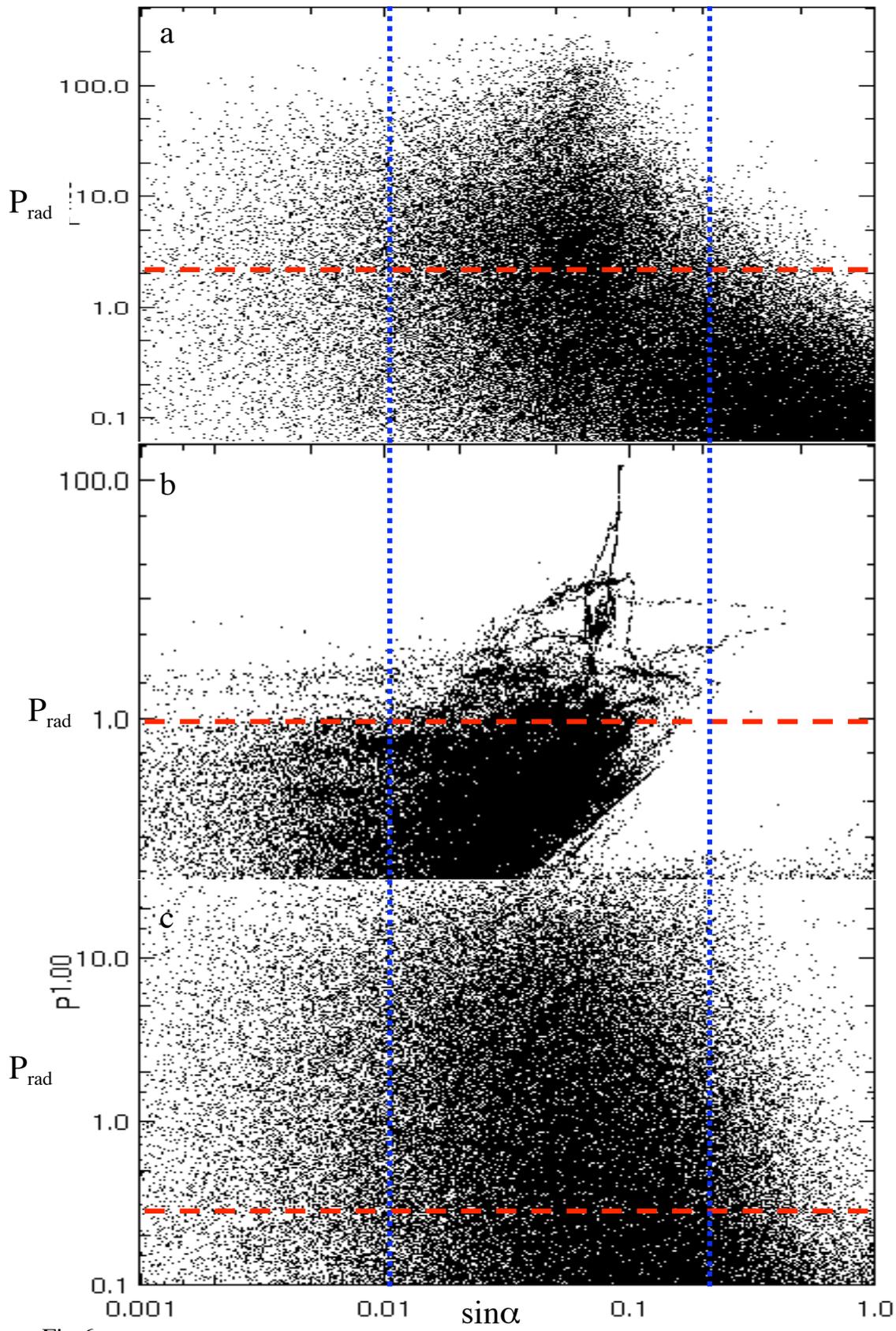

Fig.6

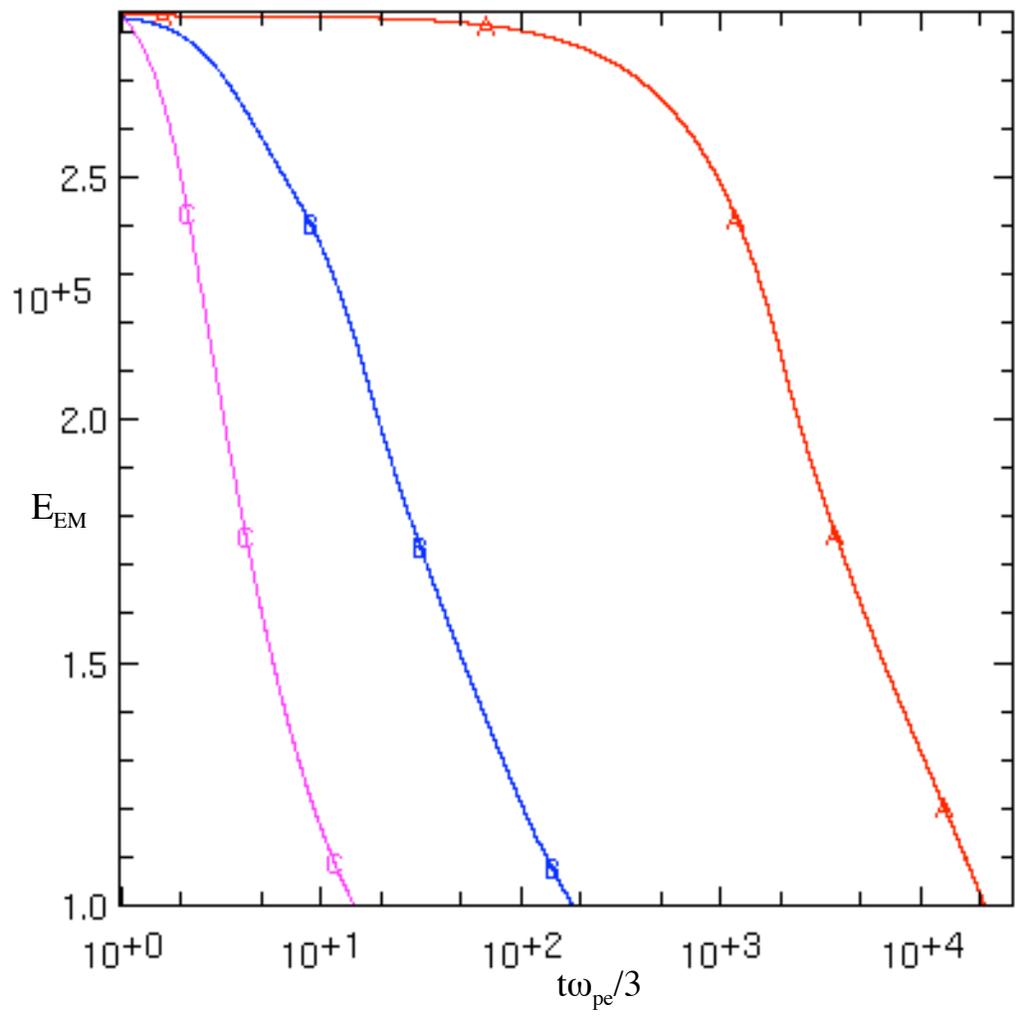

Fig.7

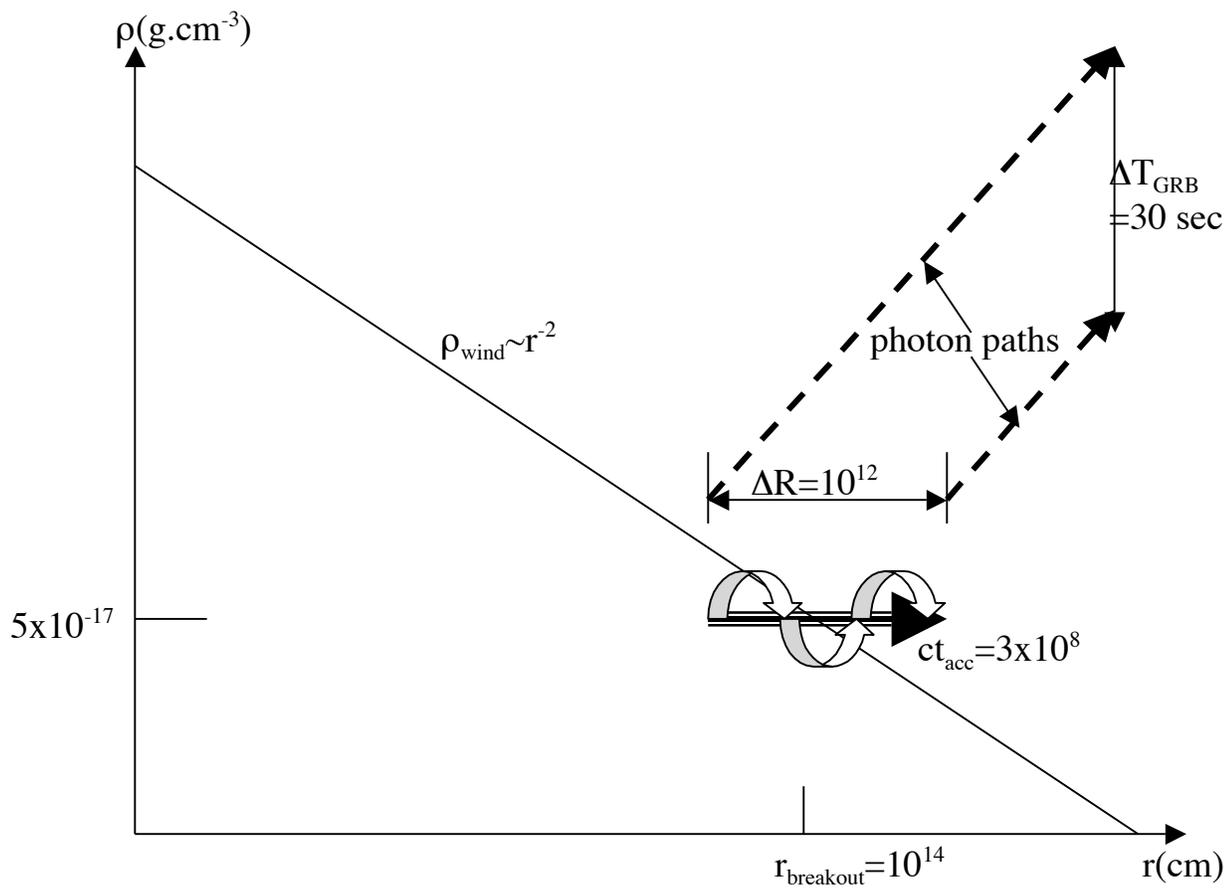

Fig.8